\documentclass[conference]{IEEEtran}

\usepackage{mathrsfs}
\usepackage{amsfonts}
\usepackage{ifpdf}
\usepackage{cite}
\usepackage{stfloats}
\ifCLASSINFOpdf
\usepackage[pdftex]{graphicx}
\else \fi
\usepackage[cmex10]{amsmath}
\usepackage{array}
\usepackage{mdwmath}
\usepackage{mdwtab}
\usepackage{eqparbox}
\usepackage[tight,footnotesize]{subfigure}
\usepackage{algorithmic}
\usepackage{algorithm}
\usepackage{amsmath}
\usepackage{epsfig}
\usepackage{ae}
\usepackage{amssymb}
\usepackage{times}
\usepackage{amsmath}   
\usepackage{booktabs}  
\usepackage{threeparttable} 
\usepackage{multirow}       
\usepackage{xcolor}
\usepackage{paralist}
\newtheorem{theorem}{Theorem}
\newtheorem{lemma}{Lemma}

\ifCLASSINFOpdf
\else
\fi

\begin{document}

\title{Throughput Capacity of Two-Hop Relay MANETs under Finite Buffers}

\author{
Jia Liu$^{1\text{,2}}$, Min Sheng$^{1}$, {\em Member, IEEE}, Yang Xu$^{1}$, Hongguang Sun$^{1}$, Xijun Wang$^{1}$, {\em Member, IEEE} \\
 and Xiaohong Jiang$^{2}$, {\em Senior Member, IEEE} \\
\IEEEauthorblockA{$^{1}$State Key Laboratory of ISN, Xidian University, Xi'an 710071, China.\\
 Email: \{liujia, msheng\}@mail.xidian.edu.cn, yxu@xidian.edu.com, hgsun@xidian.edu.cn, xjwang22@gmail.com\\
$^{2}$School of Systems Information Science, Future University Hakodate, Japan. Email: jliu871219@gmail.com, jiang@fun.ac.jp}
}
\maketitle

\begin{abstract}
Since the seminal work of Grossglauser and Tse \cite{key-1}, the two-hop relay algorithm and its variants have been attractive for mobile ad hoc networks (MANETs) due to their simplicity and efficiency. However, most literature assumed an infinite buffer size for each node, which is obviously not applicable to a realistic MANET. In this paper, we focus on the exact throughput capacity study of two-hop relay MANETs under the practical finite relay buffer scenario. The arrival process and departure process of the relay queue are fully characterized, and an ergodic Markov chain-based framework is also provided. With this framework, we obtain the limiting distribution of the relay queue and derive the throughput capacity under any relay buffer size. Extensive simulation results are provided to validate our theoretical framework and explore the relationship among the throughput capacity, the relay buffer size and the number of nodes.
\end{abstract}


\IEEEpeerreviewmaketitle

\section{Introduction}

The self-autonomous and inherent highly dynamic characteristics of mobile ad hoc networks (MANETs) make the routing schemes based on opportunistic transmission widely applied. Among them, the two-hop relay algorithm which is first proposed by Grossglauser and Tse \cite{key-1} and its variants \cite{key-2}, have become a class of attracting routing protocols due to their simplicity and efficiency \cite{key-3}. The basic idea of this protocol is that the source node can transmit packets to the destination directly, or to a node (served as a relay) it encounters, later the relay node carrying the packet forward it to the destination node. Thus, each packet travels at most two hops to reach the destination. 

By now, a lot of work has been committed to studying the throughput capacity of two-hop relay algorithm in MANETs \cite{key-1,key-2}, \cite{key-4}-\cite{key-6}. Grossglauser and Tse (2002) \cite{key-1} studied the order sense of per node capacity and showed that it is possible to achieve a $\Theta(1)$ throughput by employing the node mobility, which is a substantial improvement for the static networks considered by Gupta and Kumar (2000) \cite{key-7}. Neely \emph{et al.} (2005) \cite{key-2} computed the exact capacity and the end-to-end
queuing delay for a cell-partitioned MANET, and a fundamental tradeoff between throughput capacity and delay has been developed. Sharma \emph{et al.} (2006) \cite{key-4} proposed a global perspective for delay and capacity tradeoff in MANETs. They considered a general mobility model and related the nature of delay-capacity tradeoff to the nature of node motion. The throughput capacity with packet redundancy has been researched in \cite{key-5}, while the throughput capacity under power control has been examined in \cite{key-6}. For a detailed survey, please see \cite{key-5} and the reference therein. 

It is notable that all available work mentioned above assumed the node buffer is infinite. However, this assumption will not hold for a realistic MANET obviously. In \cite{key-8}, J. Herdtner and E. Chong have studied the throughput and storage tradeoff, and they showed that the limited relay buffer will degrade the throughput capacity. Since they only provided a scaling law relationship, the exact throughput capacity of two-hop relay MANETs under finite relay buffer size remains largely unknown by now.

As a first step towards this end, in this paper, we analytically study the exact throughput capacity of two-hop relay MANETs with the consideration that the relay buffer of each node, which is used for storing other nodes' packets, is strictly bounded. The main contributions of this paper are summarized as follows.

\begin{itemize}
\item Considering the soure-to-relay transmission under finite relay buffer scenario, we carefully compute the arrival rate at the relay queue. By utilizing the \emph{occupancy probability} technique, we exactly characterize the departure process of the relay queue.

\item Based on the queuing process of the relay node, a finite-state \emph{ergodic Markov chain} model is constructed to obtain the limiting distribution of the relay queue. With this framework, we proceed to derive the
exact throughput capacity under any relay buffer size.

\item Extensive simulation results are provided to validate our new analysis model and explore how the throughput capacity varies with the relay buffer size and the number of nodes. The results indicate that the
throughput capacity under finite relay buffer cannot stay constant as the network size growing, which is quite different from the infinite buffer scenario. 
\end{itemize}

The remainder of this paper is outlined as follows. The system models and a modified two-hop relay scheme are introduced in Section~\ref{section:preliminaries}. In Section~\ref{section:throughput}, we develop the ergodic Markov chain-based framework to fully characterize the queuing process of relay nodes and proceed to obtain the exact throughput capacity. The simulation results are provided in Section~\ref{section:simulation}. Finally, we conclude this paper in Section~\ref{section:conclusion}.

\section{Preliminaries} \label{section:preliminaries}

\subsection{System Models}

\begin{figure}[!t]
\centering
\includegraphics[width=3.0in]{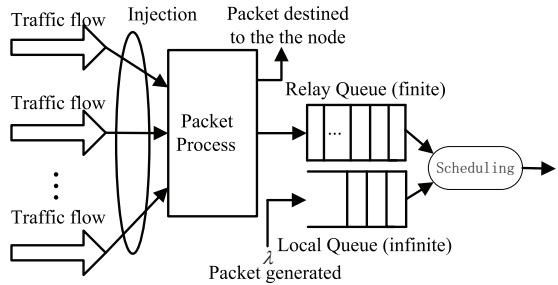} \caption{The local queue and relay queue in a node.}
\label{fig:queue_structure}
\end{figure}

\emph{Network model}: The \emph{cell partitioned} network model \cite{key-2} is adopted. The network is partitioned into $C$ non-overlapping cells of equal size. $N$ mobile nodes roam from cell to cell over the network according to the independent and identically distributed (i.i.d) mobility model \cite{key-1,key-2}. With i.i.d mobility model, at the beginning of each time slot, each node selects a cell among the network uniformly and independently, then stays in this cell during this time slot. Thus, i.i.d mobility model can be served as the limit of infinite mobility. Further, we assume that time is slotted with a fixed length, and during each time slot, only one node in each cell can transfer exactly one packet to another node in the same cell. Nodes located in different cells cannot communicate with each other. We assume each node has a local queue and a relay queue. The local queue is used to store the self-generated packets and there is no constraint on it; while the relay queue is used to store the packets from other nodes and the buffer size is set to be $B$ (packets) \cite{key-8}. The reason for this buffer assumption is elaborated in \cite{key-9}, where the ingress buffer and the internal buffer correspond to the local buffer and the relay buffer respectively. Thus, the node model can be represented in Fig.~\ref{fig:queue_structure}.

\emph{Traffic model}: We consider the traffic pattern widely adopted in previous studies \cite{key-2,key-10}, where $N$ is even and the source-destination pairs are composed as follows: $1\leftrightarrow2$, $3\leftrightarrow4$, $\cdots$, $(N-1)\leftrightarrow N$. Thus, there are in total $N$ distinct unicast traffic flows, each node is the source of a traffic flow and meanwhile the destination of another traffic flow. The exogenous packet arrival at each node is a Bernoulli process with rate $\lambda$ packets/slot.

\subsection{A Modified Two-Hop Relay Algorithm}

In this section, we make a modification on the traditional two-hop relay algorithm, to be applicable under the finite relay buffer scenario. We consider that a source node encounters a relay node and try to send a packet, if the relay queue is full, this transmission fails leading to the packet loss and the energy waste. To avoid this phenomenon, we introduce a handshake mechanism before each source-to-relay transmission to confirm the relay buffer occupancy state, if the relay queue is not full, the transmission can be conducted; else, the source node remains idle. At any time slot for each cell containing at least two nodes, the cell executes the modified two-hop relay (M2HR) algorithm which is summarized in Algorithm~\ref{algorithm:M2HR}.

\begin{algorithm}[!ht]
\caption{M2HR routing algorithm}
\label{algorithm:M2HR}
\begin{algorithmic}[1]
\IF{There exist source-destination pairs within the cell}
  \STATE With equal probability, randomly choose such a pair to do source-to-destination transmission. 
	\IF{The source contains a new packet}
	  \STATE The source transmits the new packet to the destination.
	\ELSE
	  \STATE The source remains idle.
	\ENDIF
\ELSE
  \STATE With equal probability, randomly designate a node within the cell as the sender.
	\STATE Independently choose another node among the remaining nodes within the cell as the receiver.
	\STATE Flips an unbiased coin;
	\IF{It is the head}
	  \STATE The sender conducts source-to-relay transmission with the receiver.
		\IF{The sender has a new packet, \emph{meanwhile the receiver informs that the relay queue is not full}}
		  \STATE The sender transmits that packet to the receiver.
		\ELSE
		  \STATE The sender remains idle.
		\ENDIF
	\ELSE
	  \STATE The sender conducts relay-to-destination transmission with the receiver.
    \IF{The sender has packets destined for the receiver}
      \STATE The sender transmits a packet to the receiver.
		\ELSE
		  \STATE The sender remains idle.
		\ENDIF
	\ENDIF
\ENDIF
\end{algorithmic}
\end{algorithm}

\section{Throughput Capacity Analysis} \label{section:throughput}

In this section, we first introduce some basic probabilities. Then, an \emph{irreducible ergodic} Markov chain-based framework is established to explore the occupancy distribution on the relay queue. In order to solve the Markov chain, we further investigate the departure process of the relay queue. Based on this framework, we proceed to derive the exact throughput capacity.

\subsection{Some Basic Probabilities}

For a given time slot and a particular cell, we denote by $p$ and $q$ the probability that there are at least two nodes and at least one source-destination pair in a cell, respectively. Then, the same as \cite{key-2}, we have 
\begin{align}
& p=1-\left(1-\frac{1}{C}\right)^{N}-\frac{N}{C}\left(1-\frac{1}{C}\right)^{N-1}, \\
& q=1-\left(1-\frac{1}{C^{2}}\right)^{N/2}.
\end{align}

Under the M2HR algorithm, we denote by $p_{sd}$, $p_{sr}$ and $p_{rd}$ the probabilities that a given node has a chance to conduct a source-to-destination transmission, source-to-relay transmission, and relay-to-destination
transmission at a given time slot, respectively. Then we have
\begin{align}
& p_{sd}=\frac{C}{N}q, \\
& p_{sr}=p_{rd}=\frac{C(p-q)}{2N}.
\end{align}
The derivations of probabilities $p_{sd}$, $p_{sr}$ and $p_{rd}$ are omitted here due to space limit, and please kindly refer to Appendix B in \cite{key-2} for details.

\subsection{Analysis of the Local Queue and the Relay Queue}

Under the M2HR algorithm, each packet experiences at most two queuing processes, i.e., the packet dispatching process at the local queue and the packet forwarding process at the relay queue (if the packet cannot be transmitted to the destination directly). 

\emph{Local Queue}: Due to the i.i.d mobility, the local queue can be represented as a Bernoulli/Bernoulli queue, where in every time slot a new packet will arrive with probability $\lambda$, and a service opportunity will arise with a corresponding probability $\mu_S$ which is determined as
\begin{equation}
\mu_{S}(\lambda)=p_{sd}+p_{sr}\left(1-P_{B}\right), \label{eq:mu_S}
\end{equation}
where $P_B$ denotes the probability that the relay queue is full. Note that the Bernoulli/Bernoulli queue is reversible \cite{key-11}, so the output process is also a Bernoulli flow with rate $\lambda$.

\emph{Relay Queue}: Since a specific packet from the output process of a local queue is transmitted to relay nodes with probability $\frac{p_{sr}\left(1-P_{B}\right)}{\mu_S}$, each of the $N-2$ relay nodes are equally likely to encounter the source, and for each relay node there are in total $N-2$ independent output processes of local queues may arrive at its relay queue. We denote by $\tilde{\lambda}$ the packet arrival rate at a relay queue when it is not full (when a relay queue is full, its input rate is $0$), then we have
\begin{equation}
\tilde{\lambda}\cdot(1-P_{B})+0\cdot P_{B}=(N-2)\lambda\cdot\frac{p_{sr}\left(1-P_{B}\right)}{\mu_{S}(\lambda)}/(N-2), \nonumber
\end{equation}
\begin{equation}
\Rightarrow\tilde{\lambda}=\frac{\lambda p_{sr}}{\mu_{S}(\lambda)}.\label{eq:lambda_r}
\end{equation}

\begin{figure}[!t]
\centering
\includegraphics[width=3.0in]{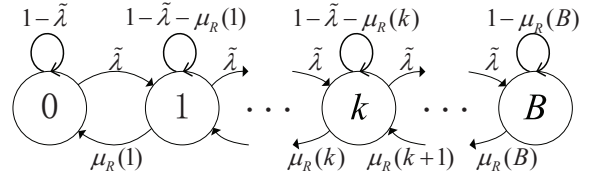}\caption{State transition diagram for the relay queue occupancy process.}
\label{fig:state_machine}
\end{figure}

We denote by $\mu_R(k)$ that the service rate when the relay queue contains $k$ packets and it is clear that $\mu_{R}(k)>0$ for $1\leq k<B$. Thus, the relay queue can be modeled as a discrete Markov chain as illustrated in Fig.~\ref{fig:state_machine} and the corresponding one-step transition matrix of the relay queue occupancy process is given by

\begin{equation}       
\mathbf{P}\!=\!\left[                 
\begin{array}{ccccc}   
\!1\!-\!\tilde{\lambda} \!&\! \tilde{\lambda} \!&\!                                    \!&\!                 \!&\!            \\  
\!\ddots                \!&\! \ddots          \!&\! \ddots                             \!&\!                 \!&\!            \\  
\!                      \!&\! \mu_R(k)        \!&\! 1\!-\!\tilde{\lambda}\!-\!\mu_R(k) \!&\! \tilde{\lambda} \!&\!            \\
\!                      \!&\!                 \!&\! \ddots                             \!&\! \ddots          \!&\! \ddots     \\
\!                      \!&\!                 \!&\!                                    \!&\! \mu_R(B)        \!&\! 1\!-\!\mu_R(B)
\end{array}
\right],           
\end{equation}

For this Markov chain we have the following lemma.

\begin{lemma} \label{lemma:ergodic}
The limiting distribution of the relay queue occupancy process exists and is unique, and is equal to the stationary distribution.
\end{lemma}

\begin{IEEEproof}
For any state $i,j\in\mathbf{S}$, $i<j$, there exist $m$ and $\tilde{m}$, such that $p_{ij}^{(m)}>0$ and $p_{ji}^{(\tilde{m})}>0$. For example, $p_{ij}^{(j-i)}=\tilde{\lambda}^{(j-i)}>0$ and $p_{ji}^{(j-i)}=\mu_{R}(j)\times\mu_{R}(j-1)\times\cdots\times\mu_{R}(i+1)>0$. Thus, any two states $i,j\in\mathbf{S}$ can \emph{communicate} with each other, which is denoted by $i\leftrightarrow j$. This type of Markov chain is called \emph{irreducible} (see, for example, Definition 4.1 in \cite{key-12} or {[}13, ch. 4, p. 168{]}), and in this Markov chain all the states are in the same class. From \cite{key-13} it follows that all the states are \emph{recurrent}. 

We denote by $d(i)$ that the period of state $i$. Note that $p_{00}^{(1)}=1-\tilde{\lambda}$, $p_{BB}^{(1)}=1-\mu_{R}(B)$, and $p_{kk}^{(1)}=1-\tilde{\lambda}-\mu_{R}(k)$, for $0<k<B$. Thus, for any state $i\in\mathbf{S}$, we have $d(i)=1$. The state $i$ and the Markov chain are called \emph{aperiodic} \cite{key-12}. Since all the states are \emph{recurrent} and \emph{aperiodic}, the relay queue occupancy process is an \emph{irreducible} \emph{ergodic} Markov chain. Referring to {[}12, ch. 5{]}, the limiting distribution of the relay queue occupancy process exists and is unique, and is equal to the stationary distribution.
\end{IEEEproof}

We use $\Pi=\left\{ \pi(0),\pi(1),\cdots\pi(B)\right\} $ to denote the limiting distribution of the relay queue. By lemma~\ref{lemma:ergodic} we have
\begin{equation*}
\Pi\cdot\mathbf{P}=\Pi,
\end{equation*}
\begin{equation}
\Rightarrow\begin{cases}
\tilde{\lambda}\pi(0)=\mu_{R}(1)\pi(1),\\
\tilde{\lambda}\pi(1)=\mu_{R}(2)\pi(2),\\
...\\
\tilde{\lambda}\pi(B-1)=\mu_{R}(B)\pi(B).
\end{cases}
\end{equation}
Combining with the normalization equation $\sum_{k=0}^{B}\pi(k)=1$, the limiting distribution of the relay queue is given by
\begin{align}
& \pi(0)=\left(1+\sum_{k=1}^{B}\frac{\tilde{\lambda}^{k}}{\mu_{R}(k)!}\right)^{-1},   \label{eq:pi_0}  \\
& \pi(k)=\frac{\tilde{\lambda}^{k}}{\mu_{R}(k)!}\left(1+\sum_{k=1}^{B}\frac{\tilde{\lambda}^{k}}{\mu_{R}(k)!}\right)^{-1}, \label{eq:pi_k}
\end{align}
where $0<k\leq B$, $\mu_{R}(k)!=\mu_{R}(k)\times\mu_{R}(k-1)\times\cdots\mu_{R}(1)$. From the (\ref{eq:pi_0}) and (\ref{eq:pi_k}), we can see that in order to derive the limiting distribution of the relay queue, we need to
compute the service rate $\mu_R(k)$ when the relay queue is on state $k$.

\subsection{Computation of the Service Rate at Relay Queue}

We denote by $p_{k}^{(i)}$ the probability that the relay queue has $k$ packets and these packets are destined for $i$ different destination nodes, $1\leq i\leq k$. Due to the i.i.d mobility model, we have
\begin{equation}
\mu_{R}(k)=\sum_{i=1}^{k}p_{k}^{(i)}\cdot i\cdot\frac{p_{rd}}{N-2}. \label{eq:mu_R_k}
\end{equation}

\begin{figure}[t]
\centering
\includegraphics[width=3.0in]{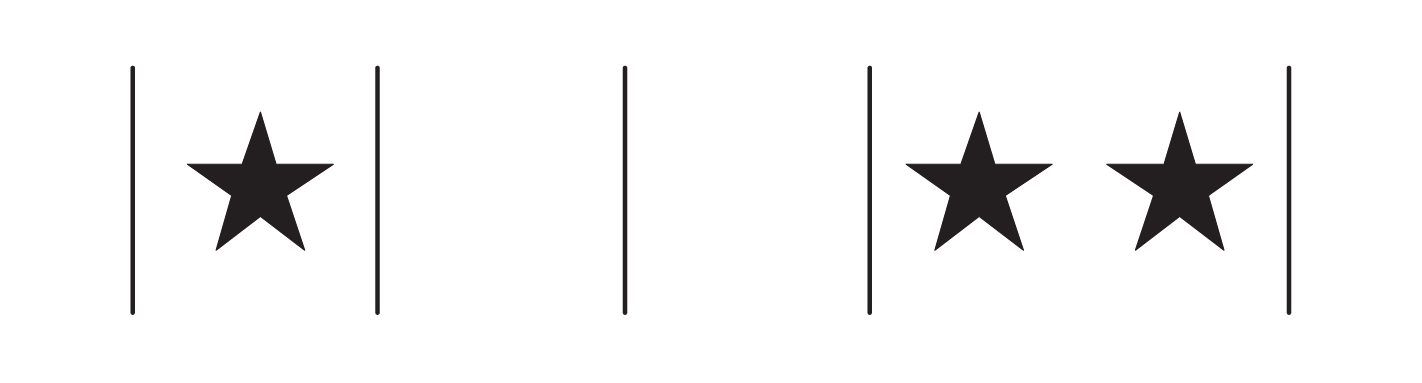}\caption{Examples of occupancy.}
\label{fig:occupancy}
\end{figure}

From the (\ref{eq:mu_R_k}), it is clear that in order to compute $\mu_{R}(k)$, we need to derive $p_{k}^{(i)}$. To address this issue, we utilize the \emph{occupancy} technique (see, for example, {[}14, ch. 1{]}). We represent the packets by 'stars' and the $N-2$ destination nodes by the spaces between $N-1$ 'bars'. For example, Fig.~\ref{fig:occupancy} represents one packet is destined for the first node, no packet is destined for the second and third nodes and two packets are destined for the fourth.

Notice that $N-2$ nodes need $N-1$ 'bars', since the first and last symbols must be 'bars', only $N-3$ 'bars' and $k$ 'stars' can appear in any order. The number of all possible permutations $E_{k}$ is given by
\begin{equation}
E_{k}=\binom{N-3+k}{k}.
\end{equation}
Considering these packets are destined for $i$ different destination nodes. The number of permutations $E_{k}^{(i)}$ is given by
\begin{equation}
E_{k}^{(i)}=\binom{N-2}{i}\cdot \binom{i-1+k-i}{k-i}.
\end{equation}
According to the classical probability we have
\begin{equation}
p_k^{(i)}=\frac{E_{k}^{(i)}}{E_{k}}=\frac{\binom{N-2}{i} \cdot \binom{k-1}{k-i}}{\binom{N-3+k}{k}}.\label{eq:p_k^i}
\end{equation}
Substituting (\ref{eq:p_k^i}) into (\ref{eq:mu_R_k}) we have
\begin{equation}
\mu_R(k)=p_{rd}\cdot\left\{ \sum\limits_{i=1}^k  \frac{\binom{N-2}{i} \cdot \binom{k-1}{k-i}}{\binom{N-3+k}{k}} \cdot \frac{i}{N-2}\right\} .\label{eq:mu_r_k_1}
\end{equation}

\subsection{Throughput Capacity}

\emph{Definition: Throughput Capacity}: For a MANET under M2HR and with a given packet arrival rate $\lambda$, the network is called stable if the queue length in each node (thus the average delay) is bounded. The throughput capacity $T_{c}$ of the network is then defined as the maximum value of $\lambda$ the network can stably support. 

\begin{theorem}
For a MANET with finite relay buffers, the throughput capacity under M2HR satisfies
\begin{equation}
T_{c}=max\{\lambda:\lambda<\mu_{S}(\lambda)\}.
\end{equation}
\end{theorem}

\begin{IEEEproof}
It is notable that $P_{B}=\pi(B)$ and this equation contains the single unknown quantity $P_{B}$. Thus, given any packet arrival rate $\lambda$, by solving the equation we can obtain $P_{B}$, and proceed to derive $\mu_{S}(\lambda)$ by (\ref{eq:mu_S}). 

Since the relay buffer size is strictly bounded by $B$ (packets), then no matter what the input rate $\lambda$ is, the relay queue is always stable. For $\lambda\in\{\lambda:\lambda<\mu_{S}(\lambda)\}$, the average delay in the local queue is given by \cite{key-11}
\begin{equation}
E\left[D_{S}\right]=\frac{1-\lambda}{\mu_{S}(\lambda)-\lambda}<\infty,
\end{equation}
thus the network is stable. When $\lambda\notin\{\lambda:\lambda<\mu_{S}(\lambda)\}$, for the Bernoulli/Bernoulli queue, the queue length will tend to infinity. Thus, the network cannot support the input rate stably. Since $T_{c}=max\{\lambda:\lambda<\mu_{S}(\lambda)\}$, then $T_{c}$ is the throughput capacity of the network.
\end{IEEEproof}

\section{Simulation} \label{section:simulation}

In this section, we first provide the simulation results to validate our theoretical framework, and then we proceed to explore how the throughput capacity be influenced by the network parameters.

\subsection{Simulation Setting}

For model validation, a simulator in C++ was developed to simulate the packet delivery process in the concerned MANET. We focus on a specific node and count its received packets over a period of $2\times10^{8}$ time slots, to calculate the time-averaged throughput. Besides the i.i.d mobility model, another realistic mobility model, the random walk model was also implemented in the simulator. With random walk mobility model, at the beginning of each time slot, each node selects a cell among its current cell and its $8$ adjacent cells with equal probability, then stays in it during this time slot.

\subsection{Model Validation}

\begin{figure}[!t]
\centering
\includegraphics[width=3in]{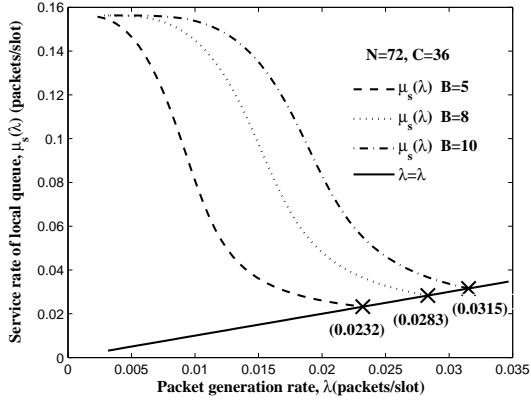}\caption{Service rate $\mu_{S}$ of the local queue vs. packet generation rate $\lambda$.}
\label{fig:lambda_vs_mu}
\end{figure}

\begin{figure}[!t]
\centering\includegraphics[width=3in]{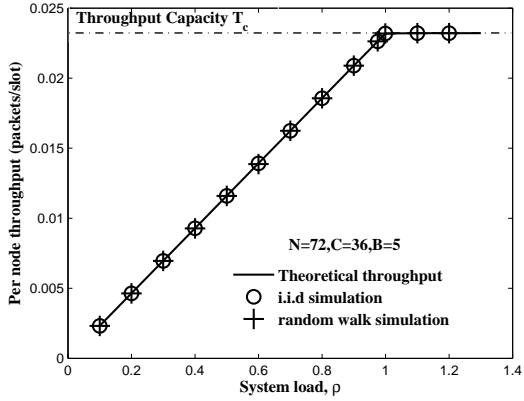}\caption{Per node's throughput performance.}
\label{fig:throughput}
\end{figure}

We fix the node number $N=72$, the cell number $C=36$, and consider three cases of $B=5$, $B=8$ and $B=10$. We increase the packet generation rate $\lambda$ step by step, and calculate the service rate $\mu_{S}$ based on the analytical framework provided. The corresponding results are summarized in Fig.~\ref{fig:lambda_vs_mu}. We can see that for all the three cases there, as $\lambda$ increasing, the service rate $\mu_{S}$
monotonically decreases (note that in the infinite buffer scenario, the service rate of local queue has no relationship with $\lambda$ \cite{key-2,key-15}), and the curves $\mu_{S}=\mu_{S}(\lambda)$ and $\lambda=\lambda$ intersect with each other. The intersection points are the exact values of throughput capacity for $B=5$, $B=8$
and $B=10$, respectively. The results are $T_{c}|_{B=5}=0.0232$, $T_{c}|_{B=8}=0.0283$ and $T_{c}|_{B=10}=0.0315$, it indicates that the larger relay buffer size leads to a higher throughput capacity.

To validate our theoretical results, we summarize the simulated throughput performance of network scenario $(N=72,C=36,B=5)$ in Fig.~\ref{fig:throughput}, where the throughput capacity $T_{c}$ is obtained by Fig.~\ref{fig:lambda_vs_mu} and the system load $\rho$ is defined as the ratio between $\lambda$ and $T_{c}$. We can see that the simulated throughput linearly increases with $\rho$ until $\rho=1$, when $\rho>1$, the throughput no longer grows up and stays as a constant which is consistent with our theoretical analysis. It is interesting to observe from Fig.~\ref{fig:throughput} that, the performance behavior under random walk mobility model is very similar to that under i.i.d mobility model. As shown in \cite{key-2}, for a cell-partitioned MANET, the throughput capacity under i.i.d model is identical to those under non-i.i.d models, if these models follow the same-steady distribution.

\subsection{Performance Analysis}

\begin{figure}[!t]
\centering
\includegraphics[width=3in]{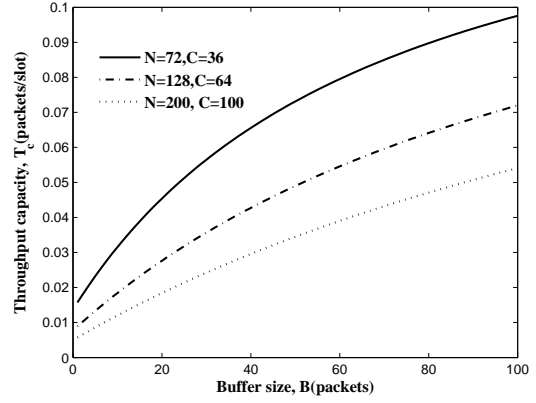}\caption{Throughput capacity $T_{c}$ vs. relay buffer size $B$.}
\label{fig:th_vs_B}
\end{figure}

\begin{figure}[!t]
\centering
\includegraphics[width=3in]{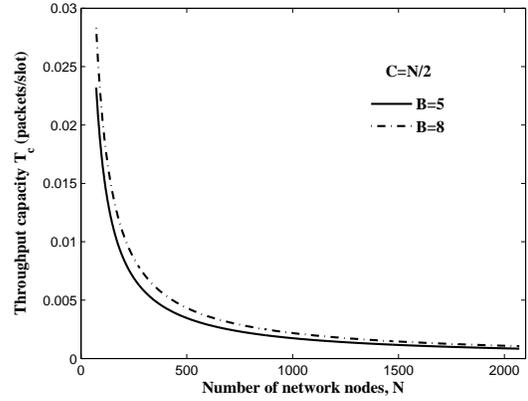}\caption{Throughput capacity $T_{c}$ vs. number of nodes $N$.}
\label{fig:th_vs_N}
\end{figure}

Fig.~\ref{fig:th_vs_B} shows the relationship between the throughput capacity $T_{c}$ and the relay buffer size $B$, under different settings of node number. The ratio between $N$ and $C$ is fixed by 2. From Fig.~\ref{fig:th_vs_B}, we can see that for all cases, $T_{c}$ monotonically increases with $B$, which indicates that the MANET really requires a sufficient relay buffer size to guarantee the throughput performance. A further careful observation is that when the buffer size is large, the increasing of the throughput capacity becomes smooth. It indicates that when the buffer size is large enough, if we continue to increase it, the performance gain will be little. It provides a guideline for the practical applications that it is effective to choose an appropriate buffer size to ensure the network performance as well as save the cost.

We proceed to explore the relationship between $T_{c}$ and $N$, where $N/C$ is fixed by 2. The corresponding results are represented in Fig.~\ref{fig:th_vs_N}. We can see that for both of the two cases there, $T_{c}$ monotonically decreases with $N$ increasing, and vanishes to 0 as $N$ tends to infinity. It is notable that the results are quite different from the throughput capacity under infinite buffer scenario, where $T_{c}$ can keep constant (about 0.14 packets/slot) with the network size increasing \cite{key-1,key-2}. It indicates that the throughput capacity can not be increased by utilizing the node mobility alone, it also requires that the relay buffer of each node increases with the network size.

\section{Conclusions} \label{section:conclusion}

In this paper, we focus on the throughput capacity of MANETs under finite buffer scenario. A modified two-hop relay routing algorithm has been proposed. We have provided an ergodic Markov chain-based framework to fully characterize the queuing process of relay nodes, and further derived the exact throughput capacity. Extensive simulation results have been conducted to verify the efficiency of the new theoretical framework and show the relationship between $T_{c}$ and $B$, $N$. The results indicate that the throughput capacity can not stay constant under the finite relay buffer scenario and increasing the relay buffer size can upgrade the throughput capacity.

\end{document}